\documentclass{sig-alternate-05-2015}

\usepackage[
pass,
]{geometry}

\usepackage{ifxetex}

\ifxetex
\usepackage{fontspec}
\else
\usepackage[T1]{fontenc}
\usepackage[utf8]{inputenc}
\fi

\usepackage{microtype}
\usepackage[hidelinks]{hyperref}

\usepackage{csquotes}

\begin{document}
\setcopyright{rightsretained}

\CopyrightYear{2016} 

\doi{http://dx.doi.org/10.1145/2897586.2897610}

\isbn{978-1-4503-4155-4/16/05}

\conferenceinfo{ICSE '16}{May 16-16 2016, Austin, TX, USA}

\acmPrice{\$15.00}

%

\title{Are easily usable security libraries possible and how should experts work together to create them?}

%
%
%
%
%

\numberofauthors{1} 
%
\author{
%
%
\alignauthor
Kai Mindermann\\
       \affaddr{University of Stuttgart}\\
       \affaddr{Institute of Software Technology}\\
       \email{kai.mindermann@informatik.uni-stuttgart.de}
}
\date{2016-01-13}

\maketitle
\begin{abstract}
Due to non-experts also developing security relevant applications it is necessary to support them too. Some improvements in the current research may not reach or impact these developers. Nonetheless these developers use security libraries. There are findings that even their usage is not easily possible and applications are left vulnerable to supposedly treated threats. So it is important to improve the usability of the security libraries. This is itself is not straightforward because of a required maturing process for example. By getting together experts of different involved areas, especially cryptographic and API-usability experts, both of the problems can be tackled.
\end{abstract}

\ccsdesc[500]{Security and privacy~Usability in security and privacy}
\ccsdesc[300]{Security and privacy~Software security engineering}
\ccsdesc[100]{Software and its engineering~Software libraries and repositories}
\ccsdesc[100]{Software and its engineering~Software evolution}

\printccsdesc


\keywords{Abstraction, API, developer knowledge}

\section{Introduction}
The security oriented-branch of the software engineering community, proposes a continuous flow of new tools and ideas to improve the overall security of developed applications and the software lifecycle. The ideas and tools have very different approaches on how they want to improve security. 
There are many tools to analyze the security during or after the development and there are many tools and ideas to model threats and risks. But is security really improved through that?

I think we have to remember that there are software developers that are unexperienced and/or non-security experts but still develop security-relevant applications, maybe even without knowing it. The ideas may have low impact for those developers. This is not unintentional because it is not expected and desired that every software developer knows about every concept and tool that is proposed. Nonetheless those developers can produce relevant applications and should be supported.

Besides the separate tools, developers can be supported through security libraries which they can utilize to develop faster and more secure applications. But still many security libraries are not completely comprehensible for the semi-professional developers and make it hard for them to apply and implement them securely \cite{Fahl:2013:RSD:2508859.2516655} and give them false hope to implement secure software which in the end is not secure or not secure enough.

This is why I propose to shift the focus to improve and develop easily usable security and cryptographic libraries.

\section{Related Work}
During the analysis of different Secure Sockets Layer (SSL) Man-in-the-middle vulnerability causes, work by Fahl et al. revealed for example that \enquote{[\dots]broken SSL code was added because developers had difficulties understanding the problem[\dots]} \cite{Fahl:2013:RSD:2508859.2516655} and \enquote{[\dots]there are developers, who, while being technically adept enough to use Wireshark to check if their app's traffic is really encrypted, do not understand the nature of the threat and thus take no precautions to counter it.} \cite{Fahl:2013:RSD:2508859.2516655}. 
Building upon that Georgiev et al. argue that \enquote{the root causes of [\dots] vulnerabilities are badly designed APIs of SSL implementations[\dots]} \cite{Georgiev:2012:MDC:2382196.2382204}.

There is research that \enquote{concentrates on the usability and security of non-security-related APIs [\dots]} \cite{Myers:2016:IAU:2896587} which explicitly does not focus on security libraries \cite{Weber:2016:EEOAUAS}. 

\section{Proposition}
I base the proposed shift to improve and develop easily usable security libraries on the following assumptions: (1) Most software developers are no security experts and they have no thorough understanding of possible attack vectors or ways to exploit their software systems. (2) There are applications that are developed by those non-security-experts but their applications have a security-relevant impact for their users. (3) It is really hard to implement security concepts the right way. (4) And even by using existing security libraries it stays hard because they are not easily usable. (5) Security of applications can be provided and can be improved by using security libraries.
These assumptions seem to be true for at least some developers \cite{Fahl:2013:RSD:2508859.2516655}, applications and libraries \cite{Georgiev:2012:MDC:2382196.2382204}.

Also based on these assumptions I expect that the security of developed applications will be far better if the security libraries are easily usable and that there will be more applications that are secure.
The difference I believe to exist is that easily usable security libraries are less prone to erroneous implementation and therefore less subject to introducing vulnerabilities in the application.

This seems to be obvious and conclusive but there are a few problems that stop us from applying it to the known libraries and security concepts.

\section{Problems}

\subsection{Maturing of Security Libraries}
Recommendable security usually follows Schneier's law: \enquote{Anyone, from the most clueless amateur to the best cryptographer, can create an algorithm that he himself can't break. It's not even hard. What is hard is creating an algorithm that no one else can break, even after years of analysis. And the only way to prove that is to subject the algorithm to years of analysis by the best cryptographers around.} \cite{Schneier:2000:SCB:341000.341009}.

This is also applicable to security libraries. 
Imagine this: You or even a group of people come up with a new library which makes it really easy to use some encryption and/or signing algorithms in a programming language. You can tell everyone that your library solves all usability problems and is as secure as the existing libraries. Even if this is true, security experts won't recommend using your library because it can not be said to be secure right at the beginning. It takes a very long time for libraries in general to mature to a recommendable and usable state. I assume this time is even longer for security libraries because they should undergo extensive cryptanalysis before they should be used in applications available for end users.

This is why one can not simply present a new cryptographic library. A more practical and more beneficial way would be to enhance existing libraries to a better usable state.

\subsection{Breaking of Compatibility}
While correcting problems in the cryptographic implementation of the security library may be without consequences for the application, improving the usability of the library usually involves changing established concepts as well as changing interface functions.

Changing interface functions can lead to breaking existing application code. And that leaves the application using the now old library and it leads to having to support more versions of your library.
Changing the concepts of the library can effectively mean it is a fork and thereby a kind of new library. That would imply it is subject to the mentioned required maturing process of a security library.

\subsection{Application of Usability Research to Security Libraries}
Even if the two problems of breaking compatibility and the needed maturing of the library can be managed, it is still unknown what easily usable means for a security library and it is unknown if a security library can be easy to use and at the same time be secure.

The most part that is in fact used by developers is the API that is offered by the library. For this part there is ongoing general research to improve APIs usability \cite{Myers:2016:IAU:2896587}.

I don't think all the results of that research can be applied easily to security APIs and security libraries because making something easier comprises abstraction: There are a lot of different and very complex security algorithms and concepts available and in use. It would be very negligent to let non-security-experts decide which aspects can or should be abstracted to make the interface of the library easier to use. On the other hand increased security can lower usability \cite{Myers:2016:IAU:2896587}.

So both the library-/API-experts and the security experts are needed, to improve the usability of security libraries.

\subsection{Developer Knowledge}
An additional problem resulting from the abstraction problem is that it is not known how much abstraction is needed. 
It is explicitly not known how much the developers must know about the security concepts that they want or have to use. This heavily depends also on the used algorithms and their intuitiveness or knowledge about them.
The result could be the decision about which parameters in an API need to be hidden from the developer through appropriate default values.

Developers knowledge can influence the security of created end-user applications \cite{Fahl:2013:RSD:2508859.2516655}. It is important to think about that during the creation or modification of security libraries.

So the audience, developers using the security libraries, is needed for the work on the libraries too.


\section{Conclusions}
Concluding I think that easily usable security libraries (APIs) are possible to a certain degree. It depends on the balance between the knowledge and comprehension capabilities of the developers and the comprehensibility and complexity of the security concepts. The problem that remains is that it is insufficient if research branches work on their own on improvements to security libraries because of the mentioned conditions in the security discipline.
A development team, consisting of usability-, cryptographic- and software-library-experts, which applies the results to security libraries together, would be ideal. The integration of the cryptographic experts in the improvement process can also lead to a shortened maturing process.


%
\bibliographystyle{abbrv}
\bibliography{mybib}  
%
%

\end{document}